\documentclass[useAMS,usenatbib,usegraphicx]{mn2e}
\usepackage{times}

\newcommand{\simlt}{\lower.5ex\hbox{$\; \buildrel < \over \sim \;$}}
\newcommand{\simgt}{\lower.5ex\hbox{$\; \buildrel > \over \sim \;$}}
\newcommand{\be}{\begin{equation}}
\newcommand{\ba}{\begin{eqnarray}}
\newcommand{\ee}{\end{equation}}
\newcommand{\ea}{\end{eqnarray}}
\newcommand{\ceff}{C_{\rm eff}}
\newcommand{\bout}{B_{\rm out}}

\title[Infall versus Star Formation Efficiency]
	{On breaking the age-metallicity degeneracy in early-type galaxies: 
	  infall versus star formation efficiency}
\author[I.~Ferreras and J.~Silk]{Ignacio Ferreras and 
Joseph Silk\thanks{ferreras,silk@astro.ox.ac.uk}\\
Physics Dept. Denys Wilkinson Building, Keble Road, Oxford OX1 3RH}
\begin{document}

\date{Draft version \today}

\pagerange{\pageref{firstpage}--\pageref{lastpage}} \pubyear{2002}

\maketitle

\label{firstpage}

\begin{abstract}
The correlation between [Mg/Fe] and galaxy mass found in elliptical
galaxies sets a strong constraint on the duration of star formation. 
Furthermore, the colour-magnitude relation restricts the range of
ages and metallicities of the stellar populations. We combine 
these two constraints with a model of star formation and chemical 
enrichment including infall and outflow of gas to find that the 
trend towards supersolar [Mg/Fe] in massive ellipticals excludes a pure
metallicity sequence as an explanation of the colour-magnitude relation.
An age spread is required, attributable either to a range of star 
formation efficiencies ($\ceff$) or to a range of infall 
timescales ($\tau_f$). We find that the inferred range of stellar 
ages is compatible with the small scatter and the redshift
evolution of the colour-magnitude relation. Two alternative scenarios
can explain the data: a fixed $\tau_f$ with a mass-dependent
efficiency: $\ceff\propto M$, or a fixed $\ceff$ with mass-dependent
infall: $\tau_f\propto 1/\sqrt{M}$. We conclude that the actual
scenario may well involve a combination of these two parameters, with
mass dependencies which should span the range of those given above.
\end{abstract}


\section{Introduction}
In the ongoing quest for the Holy Grail of galaxy formation, suitable
observables are sought to uniquely and unambiguously determine the star 
formation history of galaxies. Integrated broadband photometry targeting
special regions of the spectral energy distribution such as the
4000\AA\  break allows us to make a very good initial guess. 
The colour-magnitude relation observed in early-type galaxies 
(Faber 1973; Terlevich, Caldwell \& Bower 2001) 
and its passive evolution with redshift (Stanford, 
Eisenhardt \& Dickison 1998) is  an example. The observed colours
are compatible with very old stellar populations with a correlation
between metallicity and galaxy mass which can be motivated by 
supernova-driven winds whose efficiency is modulated by the 
gravitational potential well (Larson 1974; Dekel \& Silk 1986; 
Arimoto \& Yoshii 1987; Matteucci \& Tornamb\'e 1987). 
However, the conclusions that can be drawn 
from integrated colours alone are blurred by the age-metallicity degeneracy 
(Worthey 1994). The effect of this degeneracy on the estimates of 
stellar ages was analyzed by Ferreras, Charlot \& Silk (1999).
By using the photometry of a sample of clusters over a wide redshift range 
($z\simlt 1$), they found that significant late star formation 
-- especially in intermediate and low-mass elliptical galaxies -- is 
compatible with the data. 

\begin{figure}
\includegraphics[width=3.2in]{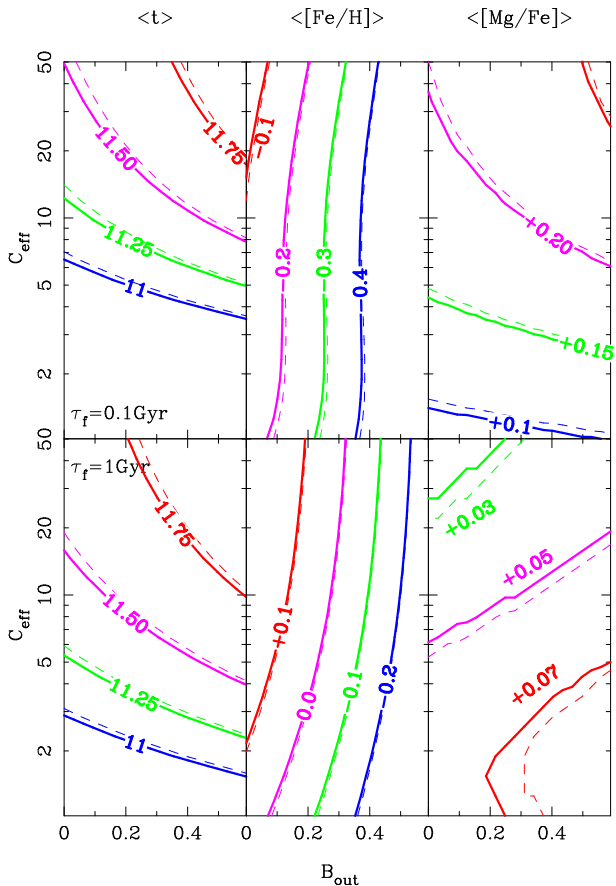}
\caption{Mass-weighted averages of stellar age ({\sl left}),
metallicity ({\sl center}), and [Mg/Fe] ({\sl right}), 
for a model with fixed formation epoch ($z_F=3$) and for two
different infall timescales: $\tau_f=0.1$~Gyr ({\sl top}) 
and $0.5$~Gyr ({\sl bottom}). The model is explored as a function
of star formation efficiency ($\ceff$) and outflow fraction ($\bout$).
The solid (dashed) contours are mass- (luminosity-) weighted values.
Notice the observed range of [Mg/Fe] cannot be reproduced 
by the model with a longer infall timescale ({\sl bottom}).
The correlations of both [Fe/H] and [Mg/Fe] with mass observed
in elliptical galaxies can be explained by either a range of efficiencies
or by a mixed $\ceff + \bout$ sequence. Notice that massive ellipticals 
require a non-negligible outflow fraction. A $\bout$ sequence (which
translates into a pure metallicity sequence) is ruled out by the 
[Mg/Fe]-mass correlation.}
\label{fig:Ceff}
\end{figure}

The failure of broadband photometry as an accurate indicator of the 
star formation history has directed attention towards narrower spectral 
windows that could help to break the degeneracy.
A combined analysis of Balmer and iron spectral indices 
from the Lick/IDS system (Worthey et al. 1994) allows us to disentangle age
and metallicity. Balmer absorption is strong in main sequence
A-type stars so that this index is especially sensitive to stars formed
in the previous $1-2$~Gyr of the galaxy history. This method quickly 
loses accuracy when older populations are considered. Furthermore, 
there are several contaminants to estimates of Balmer absorption
such as the contribution from old stars with very low-metallicities
(Maraston \& Thomas 2000) as well as  from small episodes of 
recent star formation (``frosting''; Trager et al. 2000b) 
which may introduce significant offsets between the observed 
(luminosity-weighted) and the more theoretically appealing 
mass-weighted ages.

In addition to the colour-magnitude relation which results in 
a scaling between mass and [Fe/H], elliptical galaxies display
a correlation between [Mg/Fe] and galaxy mass, so that
massive early-type galaxies are enhanced with respect to the 
predictions from population synthesis models with solar
abundance ratios (Worthey, Faber \& Gonz\'alez 1992; Davies, Sadler
\& Peletier 1993; Kuntscher 2000; Trager et al. 2000a). The main
contributors of magnesium and iron to the ISM: type~II and
type~Ia supernovae, respectively, take place on remarkably
different timescales (e.g. Matteucci \& Recchi 2001). Hence, 
[Mg/Fe] represents a valuable tool to measure star formation
timescales. Thomas, Greggio \& Bender (1999) explored this correlation
to find that formation timescales should not exceed 1~Gyr. Furthermore,
[Mg/Fe] measurements pose strong constraints on hierarchical 
clustering models, so that the baryon physics controlling 
star formation in galaxies must be significantly decoupled from the 
evolution of their dark matter halos.

The purpose of this paper is to make a combined analysis of 
broadband photometry (which is most sensitive to age and metallicity)
and the [Mg/Fe] abundance ratio (which is sensitive to the duration 
of the bursting stages) in order to constrain the star formation 
history. One aspect is crucial to this paper: to determine the possible
correlation of the various parameters that control star formation 
with a global property of the galaxy such as its total mass. We focus
on infall and outflow rates as well as on the star formation efficiency.
A previous analysis using broadband photometry 
Ferreras \& Silk (2000, paper I) concluded that there was still a degeneracy
so that both outflow rates and star formation efficiencies could
scale with galaxy mass. In this paper, we show that the addition
of [Mg/Fe] to the analysis enables us to constrain the parameter space, 
ruling out pure metallicity sequences, and allowing either the star 
formation efficiency or the infall timescale to depend on galaxy mass.

\begin{figure}
\includegraphics[width=3.2in]{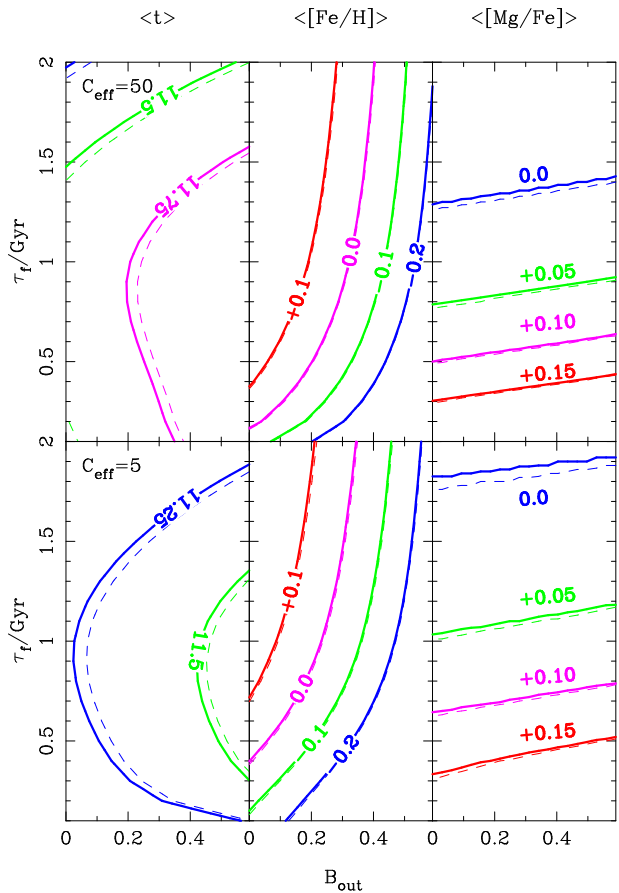}
\caption{Same as figure~\ref{fig:Ceff} for a range of infall timescales
($\tau_f$) and outflow gas fraction ($\bout$) for two star formation
efficiencies: $\ceff=100$ ({\sl top}) and $10$ ({\sl bottom}). In this
case, no trajectory in the $(\tau_f,\bout )$ can be found that
explains {\sl both} trends of [Fe/H] and [Mg/Fe] with galaxy mass. This
implies star formation efficiency and outflows are the parameters
strongly dependent on mass, whereas infall timescale should have
a weaker dependence.}
\label{fig:Tf}
\end{figure}

\section{A simple model of star formation}
We explore a one-zone model of star formation and chemical enrichment
as described in Ferreras \& Silk (2000). Each star formation
history is determined by a set of four parameters:
star formation efficiency ($\ceff$); fraction of gas and metals
ejected in outflows ($\bout$), infall timescale ($\tau_f$)
and formation epoch ($z_F$). We model the infall rate of primordial gas
by a gaussian function whose spread is given by $\tau_f$ and whose
epoch at maximum infall rate is given by the formation 
redshift $z_F$. We assume a Salpeter IMF in the mass range 
$0.1<M/M\odot<60$. The model tracks the stellar, gas and metal components. 
The yields from type~II supernovae (SNII) are taken from Thielemann, 
Nomoto \& Hashimoto (1996) for a range of progenitor masses. 
Type~Ia supernovae (SNIa) are included using the prescription of 
Greggio \& Renzini (1983) assuming a close  binary composed of 
a white dwarf and a non-denegenerate companion in the (binary) mass
range $3-16 M_\odot$. We refer the reader to Matteucci 
\& Recchi (2001) for a comprehensive review of estimates 
of SNIa rates. The yields are taken from model W7 in 
Iwamoto et al. (1999). Two elements are considered 
in this paper: magnesium and iron. The production of these
two elements is remarkably different between both supernova types.
A very significant amount of iron is produced in SNIa 
with respect to the $\alpha$ elements such as Mg. 
The mass contained in O-, Ne- and C-burning shells 
is too small compared with the mass in 
the Si-burning zone. This implies SNIa ejecta are dominated by
the products of complete and incomplete Si-burning, i.e. a higher
iron yield compared to the ejecta from core-collapse (type~II) 
supernovae. Most of the iron produced in the latter falls 
inside a mass-cut that collapses to form the stellar remnant.
Furthermore, the different timescale for the onset
of both supernova types makes abundance ratios such as [Mg/Fe]
very sensitive tracers of the duration of star formation.
Short-lived and intense bursts of star formation generate 
stellar populations mainly polluted by the elements contributed 
by SNII, resulting in enhanced [Mg/Fe]. On the other hand, a 
more extended and weaker star formation history will allow the
iron produced in SNIa's to contribute significantly to the
stellar metallicity, lowering the [Mg/Fe] ratio.

Hence, there is a direct -- albeit non-trivial -- correspondence 
between [Mg/Fe] and star formation duration. Similarly to analyses
of metallicites performed to infer the star formation history, the absolute
estimate of the lapse of star formation is highly dependent on the
stellar yields used. Thomas et al. (1999) find
a significant difference between estimates of the star formation 
timescale using the SNII yields
from Woosley \& Weaver (1995) and from Thielemann, Nomoto \& Hashimoto
(1996). The star formation timescale could be reduced by a factor
up to $100$ when using the yields from the former. The models of 
Thielemann et al. (1996) have a higher yield of Mg for stellar
masses $M\sim 20-25 M_\odot$. This discrepancy is claimed to be
caused by the different criterion used for convection. Furthermore, 
the models of Woosley \& Weaver (1995) generate more iron (models B,C).
Most of the Mg is produced during hydrostatic burning of 
the carbon shell. However, the mechanism for the production of $^{56}$Ni
which decays into $^{56}$Fe is strongly dependent on the highly
uncertain explosion mechanism. Throughout this paper we will use
the yields from Thielemann et al. (1996) for SNII. However, more 
work is definitely needed in this field if we want to make accurate
estimates of the star formation timescales.

\begin{figure}
\includegraphics[width=3.2in,height=3.5in]{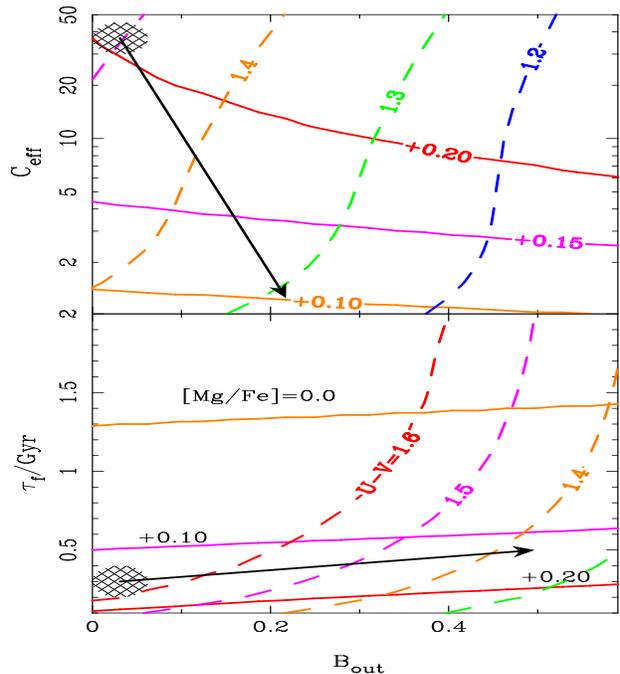}
\caption{A comparison of two alternative scenarios: the
colours and abundance ratios observed in early-type galaxies 
can be explained by a correlation between mass and either
star formation efficiencies ({\sl top}) or 
infall timescales ({\sl bottom}). The thick contours give [Mg/Fe]
abundance ratios. The prediction for a simple stellar population
with the age and metallicity determined by the model is shown 
by (dashed) contours of $U-V$ colour. The arrows give a rough locus of the
mass sequence, with the hatched region representing the most 
massive elliptical.}
\label{fig:MgFe}
\end{figure}

\section{Infall versus Efficiency}
Figure~\ref{fig:Ceff} shows mass- (solid) and luminosity-weighted (dashed) 
contours of age ({\sl left}), [Fe/H] ({\sl centre}), 
and [Mg/Fe] ({\sl right}) as a function of star formation efficiency
($\ceff$) and gas outflow fraction ($\bout$) for two infall timescales:
$\tau_f=0.1$~Gyr ({\sl top}) and $1$~Gyr. The formation epoch is $z_F=3$,
which is compatible with the best estimate of the formation redshift
of the stellar populations in ellipticals from an analysis of the
fundamental plane in clusters out to $z\sim 1.3$ (Van~Dokkum \& 
Stanford 2002).
Short infall timescales are required if we want to reproduce the
observed range of abundance ratios (e.g. Trager et al. 2000a; 
Kuntschner 2000), otherwise, choosing longer timescales would imply
very low star formation efficiencies ($\ceff < 1$) in order to 
achieve low abundance ratios [Mg/Fe]$\sim 0.0 - 0.05$, and would
generate young stellar populations which will not reproduce the 
redshift evolution of the observed slope and scatter of 
the colour-magnitude relation (Stanford et al.  1998).
The parameter range shown in figures~\ref{fig:Ceff} and \ref{fig:Tf}
is chosen to avoid such young stellar populations. The birthrate 
parameter --- $b\equiv \psi/\langle\psi\rangle$ --- for all models 
considered in this paper is below $b\simlt 0.05$, i.e. the models never reach the
lowest birthrates of early-type disks (Kennicutt, Tamblyn \& Congdon 1994).
The $\tau_f=0.1$~Gyr model shown in figure~\ref{fig:Ceff} shows 
that a $\bout$ sequence (i.e. a horizontal
line in the figure) -- in which only the outflow fraction is assumed to vary with 
galaxy mass -- could explain the mass-metallicity relation compatible
with the colour-magnitude relation, as suggested by 
Kodama \& Arimoto (1997). However, the constraint imposed by the
correlation between [Mg/Fe] and galaxy mass excludes this as a 
possibility. A range of efficiencies (``$\ceff$ sequence'', vertical 
line) or a tilted ``$\ceff +\bout$'' sequence is needed to explain 
both correlations. Hence, the combined analysis of [Mg/Fe] and [Fe/H] 
poses very powerful constraints on the star formation history.

\begin{figure}
\includegraphics[width=3.5in]{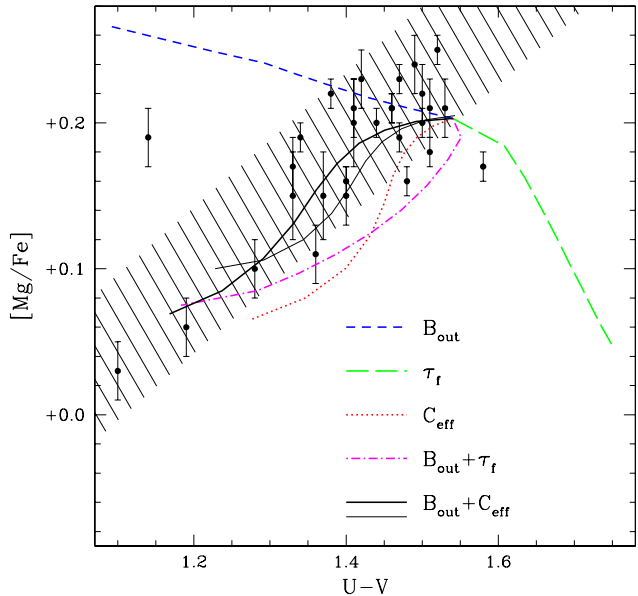}
\caption{Observed correlation between [Mg/Fe] and $U-V$ colour from
the sample of Gonz\'alez (1993) later analyzed by 
Trager et al. (2000a). Several model predictions are shown as discussed
in the text. The shaded region is a fit to the data and is used to 
determine the parameters shown in figure~\ref{fig:Mv}.}
\label{fig:Trager}
\end{figure}

Figure~\ref{fig:Tf} presents a similar plot to the previous figure.
In this case a range of infall timescales and outflow fractions are
explored. Two star formation efficiencies are considered: $\ceff =50$ 
({\sl top}) and $5$ ({\sl bottom}). The figure shows that the range 
of abundance ratios could be explained by longer infall timescales 
in low-mass galaxies. Figures~\ref{fig:Ceff} and \ref{fig:Tf}
illustrate the degeneracy between star formation efficiency
and infall timescale, as both can reproduce a similar range of star 
formation histories. However, the chemical enrichment can be used to
break this degeneracy, as seen in the central panel of 
figure~\ref{fig:Tf}. Models with long infall timescales give solar
[Mg/Fe] but they also predict higher [Fe/H]. Hence, an explanation
of the observed correlation between colour and [Mg/Fe] requires
much larger outflow fractions if a range of infall timescales is 
assumed. Figure~\ref{fig:MgFe} further illustrates this point by
overlaying contours of $U-V$ colour as predicted by population
synthesis models (Bruzual \& Charlot, in preparation) on [Mg/Fe]
contours for a model with varying star formation efficiency
({\sl top}) or infall timescale ({\sl bottom}). The hatched regions
represent the area of parameter space that best explains massive
ellipticals. The arrow gives the locus of points which track a
sequence of galaxy masses. One can see from the figure that a 
remarkably larger range of outflow fractions is needed if a
mass-independent star formation efficiency is assumed. 

\begin{figure}
\includegraphics[width=3.5in]{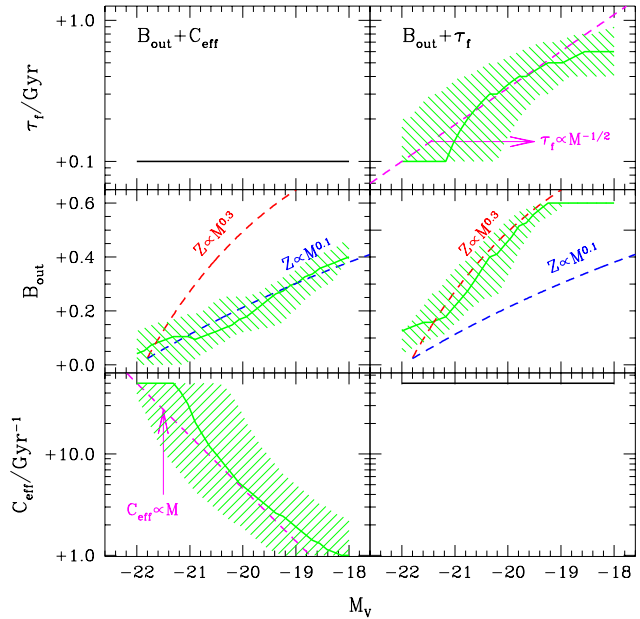}
\caption{Correlation of the model parameters with the $V$-band 
absolute luminosity, using the observed $U-V$ and [Mg/Fe] correlation
with $M_V$ as constraints. The left and right panels represent a model
with constant star formation efficiency or infall timescale, respectively.
The shaded regions in each panel are mappings of the region shown in 
figure~\ref{fig:Trager} for each parameter. The solid lines give the 
mapping of the best fit. The dashed lines are some ``educated guesses''
for the mass dependence of these parameters.}
\label{fig:Mv}
\end{figure}

Figure~\ref{fig:Trager} shows the observed correlation between
[Mg/Fe] and $U-V$ colour from the sample of Gonz\'alez (1993),
which has been analyzed in detail by Trager et al. (2000a; 2000b).
The translation from the observed spectral indices to estimates of
[Mg/Fe] is performed by correcting the population synthesis models
calibrated for solar abundance ratios using the model atmospheres
for non-solar ratios of a 5~Gyr isochrone studied by Tripicco 
\& Bell (1995). This introduces a systematic
error which prevents us from getting accurate estimates of the 
absolute value of [Mg/Fe]. However, the relative values are
more robust and the data shown in the figure gives a clear
sign of a correlation between metallicity and star formation
timescale. The lines give the model predictions when
some of the parameters are kept constant. The dashed line
is the prediction for a constant efficiency and infall timescale,
so that only the outflow fraction -- $\bout$ -- is allowed to 
vary with galaxy mass. This model is incompatible with the data
as [Mg/Fe] is predicted to increase towards blue $U-V$ colour.
The model with a varying infall timescale and constant $\ceff$
and $\bout$ (long dashed line) is also ruled out as redder
colours are predicted with decreasing [Mg/Fe]. The other three
models are -- within model and observational uncertainties -- 
compatible with the data, although the one which seems to 
give a best fit to the data is the model with a constant
infall timescale and a varying $\ceff$ and $\bout$ (solid line).
The shaded area shown in figure~\ref{fig:Trager} represents the
best fit to the data points along with a $\pm 0.05$~dex scatter:
\be
[{\rm Mg/Fe}] = 0.46(U-V)-0.46\pm 0.05.
\ee
Figure~\ref{fig:Mv} shows the mapping of this region in parameter
space for the two most plausible scenarios, namely for a fixed
infall timescale (``$\bout +\ceff$ sequence'', {\sl left}) or for a 
fixed star formation efficiency (``$\bout +\tau_f$ sequence'',{\sl right}). 
The $U-V$ colour-magnitude relation of the Coma cluster
is used in order to translate between 
colour and absolute luminosity (Terlevich et al. 2001). 
A fixed star formation efficiency
requires low-mass galaxies ($M_V>-19$) to have extremely high outflow
fractions ($\bout >0.6$). Models with a mass-dependent efficiency
require a smaller range of outflows: $0<\bout <0.4$. Doubtlessly, the
real scenario will involve a range of all three parameters considered:
$\bout$, $\ceff$ and $\tau_f$. However, the colour constraint on the 
range of outflows makes a stronger dependence of $\ceff$ 
on mass more plausible. 
The dashed lines in the panels of figure~\ref{fig:Mv} are
guesses for the correlation with mass, using the observed dependence
of $M/L_V$ on galaxy mass in cluster ellipticals (Mobasher et al. 1999). 
The infall model ({\sl top right}) gives a good fit to a 
$\tau_f\propto 1/\sqrt{M}$ scaling, which can be theoretically motivated
if we write the accretion timescale as the cross-section per unit mass:
\be
\tau_f\sim \frac{R^2}{M}\sim\frac{1}{M^{1/3}\rho^{2/3}}.
\ee
If we assume a small range of formation redshifts, the correlation
between $\tau_f$ and mass is compatible with the model predictions
shown in the top right panel of figure~\ref{fig:Mv}. On the other hand, 
the efficiency model ({\sl bottom left})
favours a linear correlation with mass: $\ceff\propto M$, a result
which can be theoretically motivated by the effects of feedback on 
star formation (Silk 2002). 
The dashed lines in the middle panels span the plausible range of 
the correlation between mass and metallicity:
$1-\bout\propto\langle Z\rangle\propto M^\alpha$, with 
$\alpha\sim 0.3-0.1$ (e.g. Larson 1974; Mould 1984; Dekel \& Silk 1986).
Hence, a shallow mass-metallicity relation ($\alpha\sim 0.1$) 
favours a model with a mass-dependent star formation efficiency, 
whereas infall timescales should be more dependent on the mass 
of the galaxy if $\alpha\sim 0.3$. All these correlations
are nevertheless strongly model-dependent, so that we could expect the real
mechanism to lie in between these two alternative scenarios, i.e. the
efficiency should be at most linearly dependent on mass and 
the infall timescale should at most vary as $1/\sqrt{M}$.

\section{Conclusions}
We have explored the correlation between $U-V$ colour 
and [Mg/Fe] in early-type galaxies to determine the role of the 
galaxy mass in the formation of this type of galaxies. Infall
timescale ($\tau_f$), star formation efficiency ($\ceff$) 
and gas outflows ($\bout$) are the
main ``drivers'' for this correlation. The formation epoch is
kept fixed throughout the paper to $z_F=3$. 
This is motivated by the fact that 
$z_F$ is tightly constrained by the observed evolution of the 
fundamental plane or the colour-magnitude relation in clusters 
out to $z\simlt 1.3$ (Stanford et al. 1998; Van~Dokkum \& Stanford 2002),
so that values $z_F\simlt 2$ are readily ruled out. Earlier 
formation epochs do not significantly change the results shown
in this paper. In the list of 
caveats that any model of chemical enrichment always carries, we 
should emphasize the strong dependence of the model predictions on 
the stellar yields. In this paper we have made use of the models of 
Thielemann et al. (1996) for type~II SNe and of Iwamoto et al. (1999) 
for type~Ia SNe. The currently available alternative for type~II SNe 
yields, namely Woosley \& Weaver (1995), gives similar metallicities, but 
rather low [Mg/Fe] abundance ratios unless extremely short infall timescales 
are considered (Thomas et al. 1999).

Pre-enrichment of the infalling gas is a mechanism which could
alter the model predictions. However, as long as we assume the abundance
ratios of the pre-enriched gas to be similar to type~II SNe ejecta, 
our model gives higher metallicities and roughly similar [Mg/Fe].
Hence, a slight increase in $\bout$ lowers [Fe/H] enough to give 
equally good fits to the data. The solid lines in figure~\ref{fig:Trager}
are ``$\bout +\ceff$'' sequences with the infalling gas having zero
metallicity (thick line) or $Z_\odot /10$ with enhanced $[$Mg/Fe$]=+0.3$~dex
(thin line). The difference between these two is negligible within the 
underlying uncertainties. This illustrates the fact that 
absolute estimates of parameters in chemical enrichment models 
are prone to large systematic errors. Nevertheless, the relative
estimates are robust and only weakly dependent on these uncertainties.

Figures~\ref{fig:Ceff} and \ref{fig:Tf} show that the [Mg/Fe]
data poses a degeneracy betweeen infall timescale and star 
formation efficiency, which is to be expected since the data can 
only constrain the duration of the star formation episode. 
The constraint on the metallicity from the colour-magnitude relation 
is added to the analysis, so that this degeneracy is partially
lifted. The model shows that a mass-dependent range of outflow
fractions is required to explain both $U-V$ and [Mg/Fe]. Furthermore,
a simple model in which {\sl only} $\bout$ scales with mass
is incompatible with the data. Hence, a pure metallicity sequence
is readily ruled out by the observations. A certain spread in the
ages is thereby expected in ellipticals, although the predicted
range of ages is for all practical purposes indistinguishable
from a fixed age population in local and moderate redshift 
ellipticals. The range of infall timescales or star formation
efficiencies has been shown to be degenerate, although 
figure~\ref{fig:Trager} might favour a scenario
in which the efficiency has a strong dependence on galaxy mass,
of order: $\ceff\propto M$. Nevertheless, a range of infall 
timescales $\tau_f\propto 1/\sqrt{M}$ is also compatible with the data and
the assumption of a mass-metallicity relation $Z\propto M^{0.3-0.1}$
favours this model. The accurate measurement of the mass-metallicity
relation will be one of the key observables for disentangling
infall timescales and star formation efficiency.


\bsp

\label{lastpage}

\end{document}